\documentclass[aps,prc,showpacs,superscriptaddress,twocolumn]{revtex4}

\usepackage{graphicx,epsfig}

\begin{document}

\title{Medium-modified Jets and Initial State Fluctuations as Sources of Charge
Correlations Measured at RHIC}

\author{Hannah Petersen}
\affiliation{Department of Physics, Duke University, Durham, North Carolina
27708-0305, United States}

\author{Thorsten Renk}
\affiliation{Department of Physics, P.O. Box 35 FI-40014 University of
Jyv\"askyl\"a, Finland}
\affiliation{Helsinki Institute of Physics, P.O. Box 64 FI-00014, University of
Helsinki, Finland}

\author{Steffen A. Bass}
\affiliation{Department of Physics, Duke University, Durham, North Carolina
27708-0305, United States}


\begin{abstract}
We investigate  the contribution of medium-modified jets and initial state
fluctuations
to the asymmetry in charged particle
production with respect to the reaction plane. This asymmetry has been suggested
as a compelling signature of the Chiral Magnetic Effect in QCD and make a study
of "conventional" scenarios for the creation of such charged particle
multiplicity
fluctuations a timely endeavor.
The different
pathlength combinations of jets through the medium in non-central heavy ion
collisions result in finite correlations of like and different charged particles
emitted in the different hemispheres. Our calculation is based on combining jet
events from YaJEM (Yet another Jet Energy Loss Model) and a bulk medium
evolution. It is found that the jet production probabilities are too small to
observe this effect. The influence of initial state fluctuations on this
observable is explored using an event-by-event (3+1)d hybrid approach that is
based on UrQMD (Ultra-relativistic Quantum Molecular Dynamics) with an ideal
hydrodynamic evolution. In this calculation momentum conservation and elliptic
flow are explicitly taken into account. The asymmetries in the initial state are
translated to a final state momentum asymmetry by the hydrodynamic flow profile.
Dependent on the size of the initial state fluctuations the resulting charged
particle asymmetries are in qualitative agreement with the preliminary STAR
results. The multi-particle correlation as proposed by the PHENIX collaboration
can in principle be used to disentangle the different contributions, however is in practice substantially 
affected by the procedure to subtract trivial resonance decay contributions.  
\end{abstract}



\maketitle

\section{Introduction}

Ultra-relativistic heavy-ion collisions at the Relativistic Heavy-Ion Collider
(RHIC)
are thought to have created a high temperature and pressure state of QCD matter
termed
the Quark-Gluon-Plasma (QGP)
\cite{Arsene:2004fa,Adcox:2004mh,Back:2004je,Adams:2005dq,Muller:2006ee}: Two
major discoveries stand out: (1) the emission of hadrons with a transverse
momentum $p_T$ 
of several GeV/c or more is strongly suppressed (jet-quenching), implying
the presence of matter 
with a very large color opacity, and (2) the anisotropic (``elliptic'') flow
in non-central collisions is near the ideal hydrodynamic limit, requiring an
early onset of the period during 
which the expansion is governed by fluid dynamics (earlier than 1 fm/c after the
initial impact) as 
well as nearly ideal fluid properties with a viscosity-to-entropy density ratio
$\eta/s \ll 1$. 
The matter created at RHIC has been thus called the strongly interacting
Quark-Gluon Plasma (sQGP) \cite{Gyulassy:2004zy} and quantifying its properties
in terms of its transport coefficients is
one of the major tasks of the current experimental and theoretical research
program.

One should note that QCD matter created in relativistic heavy-ion collisions may
possess a 
very rich set of features, reflecting the fundamental symmetries (and violations
thereof)
of QCD, well beyond the above noted characteristics of large opacity and
near-ideal fluidity.
Among the more intriguing features which are currently being searched for is the
presence of
the Chiral Magnetic Effect (CME)
\cite{Kharzeev:2007jp,Fukushima:2008xe,Fukushima:2010vw,Asakawa:2010bu}: 
as was pointed out in \cite{Kharzeev:2007jp}, non-central heavy-ion collisions 
create a coherent magnetic field that may convert topological
charge fluctuations in the QCD vacuum into global electric charge fluctuations
with respect
to the reaction plane. While the magnitude of the charge fluctuations which can
be attributed 
to the CME is still a matter of ongoing debate
\cite{Fukushima:2010vw,Asakawa:2010bu},
the STAR collaboration has recently reported that charge fluctuations
with respect to the reaction plane have indeed been seen \cite{Abelev:2009uh}.

In order to confirm the discovery of the CME in relativistic heavy-ion
collisions, one
first needs to verify that experimental measurements cannot be understood in
terms of 
"conventional", well-established, phenomena observed at RHIC. It has recently
been
pointed out that correlations from charge and momentum conservation at
freeze-out overlaid with elliptic flow may very well be able to describe the
observed phenomena \cite{Pratt:2010gy,Schlichting:2010na}. In addition,
the experimental observable suggested to measure the CME has been argued to
be predominantly sensitive to other effects \cite{Wang:2009kd,Bzdak:2009fc}.

In this paper we focus on estimating various conventional physics mechanisms that lead to charge particle number asymmetry within a realistic
model, namely medium-modified jets as well as initial state fluctuations, combined with well-known bulk physics 
such as momentum conservation, resonance decays and elliptic flow. In Section \ref{observables} the different observables that have been proposed are introduced. These observables are surely sensitive to fluctuations of multiplicity (and charge) if these fluctuations are correlated with the reaction plane. Both of the two szenarios that we are studying here, in-medium jets (in Section \ref{jets}) and initial state granularity (in Section \ref{granularity}) give a particular mechanism to generate multiplicity fluctuations correlated in a peculiar way with the reaction plane. Only once all possible contributions to the observed charged particle
fluctuations with
respect to the reaction plane are accounted for, can we quantify the possible
contribution
of the CME to the measurements.

\section{Definition of Observables}
\label{observables}

The standard observable used to look for the CME is the angular correlation coefficient

\begin{equation}
\gamma_{\alpha,\beta} = \frac{\sum_{i\in \alpha,j\in \beta}\cos(\phi_i+\phi_j)}{M_\alpha M_\beta}
\end{equation}

where $\alpha$ and $\beta$ represent positive and negative charges, $\phi$ measures the angle with respect to the reaction plane orientation and $M$ are the measured multiplicities. As argued in
\cite{Pratt:2010gy,Schlichting:2010na,Bzdak:2009fc}, this observable is (contrary to initial expectations) dominated by a strong in-plane like-sign correlation which has no obvious relation with charge separation as proposed as signal of the CME. 

In the present study we consider two other newly developed observables to quantify the correlations of charged particles with the reaction plane for which a computation using a detailed framework is feasible with a reasonable amount of CPU time. In addition, these observables seem to have a clearer relation to out-of-plane charge separation. 

\subsection{Charge Asymmetry Observable}

The first observable we focus on for measuring the event-by-event charged particle
asymmetry  has been developed by the Purdue group within the
STAR collaboration \cite{STAR_Talk}. The measurement is based on counting charged particles that
are emitted into different hemispheres with respect to the reaction plane. The
different hemispheres are defined as follows: 'Up=U' means $p_y>0$, 'Down=D'
refers to $p_y<0$, 'Left=L' stands for $p_x<0$ and 'Right=R' for $p_x>0$. The
UD-direction reflects the out-of-plane direction whereas the LR-direction
represents the in-plane direction. 

The charged particle asymmetry coefficients are defined in the following way

\begin{equation}
(A^+_{\rm UD})^2=\frac{(N^+_{\rm U}-N^+_{\rm D})^2}{(N^+_{\rm U}+N^+_{\rm D})^2}
\end{equation}

\begin{equation}
(A^-_{\rm UD})^2=\frac{(N^-_{\rm U}-N^-_{\rm D})^2}{(N^-_{\rm U}+N^-_{\rm D})^2}
\end{equation}

\begin{equation}
A^{+/-}_{\rm UD}=\frac{(N^+_{\rm U}-N^+_{\rm D})(N^-_{\rm U}-N^-_{\rm
D})}{(N^+_{\rm U}+N^+_{\rm D})(N^-_{\rm U}+N^-_{\rm D})}
\end{equation}

and correspondingly for the LR-direction where $N^+$ and $N^-$ are the numbers
of positively and negatively charged particles in one event. 

The opposite charge correlation ($A^{+/-}_{\rm UD}$ and $A^{+/-}_{\rm LR}$) are
positive, if the particles tend to be aligned and negative for a charge
separation scenario. Without any correlation this observable would be consistent
with zero. The $(A^{+})^2$ and $(A^{-})^2$ are a measure of the overall charged
particle fluctuation. The preliminary STAR results seem to be positive (on the
order of $1 \cdot 10^{-3}$) in the out-of-plane as well as the in-plane
direction for mid-central Au+Au collisions at $\sqrt{s_{\rm NN}}=200A$ GeV. The
main difference of this observable compared to the original angular correlation
measurement \cite{Abelev:2009uh,Abelev:2009txa} is that each particle
contributes with the
same weight independent of its position in the respective hemisphere. 

\subsection{Multi-Particle Correlation}
\label{sec_phenix_def}
The PHENIX collaboration has recently proposed a different observable that is
sensitive to a charge separation with respect to the reaction plane as it is
predicted from the chiral magnetic effect \cite{PHENIX_Talk}. This multi-particle correlation coefficient relies on an event-by-event measurement technique and is defined as follows
\begin{equation}
C_{p}=\frac{dN/d(\Delta S_{\rm ch})}{dN/d(\Delta S_{\rm all})} 
\end{equation}
where
\begin{equation}
\Delta S = \frac{1}{P}\sum_{i=1}^P \sin(\phi_i^+)-\frac{1}{M}\sum_{i=1}^M \sin(\phi_i^-) \quad .
\end{equation}

Here, the first sum runs over all positively charged particles in one event
whereas $M$ is the number of negatively charged particles in one event and
$\phi$ is the azimuthal angle in the transverse plane in momentum space. $\Delta
S_{\rm all}$ is the analogous quantity where the $i$th particle is randomly
chosen from the event. This observable is a multi-particle correlation that has
to be measured event by event. The ratio of the charge sensitive distribution to
the one for all charged particles is flat at 1, if there is no charge specific
correlation (e.g. elliptic flow). A concave shape indicates charge separation out of plane  (e.g. the 
presence of the CME) and a convex shape arises, if positively and negatively
charged particles are preferably emitted in the same direction (e.g. resonance
decays).

\section{Medium-modified Jets}
\label{jets}
\subsection{Overview}

The idea of jets as a mechanism to generate asymmetries in the charged
multiplicity distribution is based on the notion that in-medium energy loss from
a leading parton is dominated by induced gluon radiation
\cite{Baier:1996sk,Zakharov:1997uu}, giving rise to additional multiplicity production.
Moreover, this mechanism is predominantly active out of plane:
While radiative energy loss asymptotically has
an $L^2$ dependence in a constant medium, the actual scaling of lost energy and
produced multiplicity in a real situation is complicated and has no simple
functional shape for a number of reasons, among them the spacetime evolution of
the bulk medium which reduces the density as a function of space and time and
the re-interaction of radiated gluons with the medium, leading to a cascade of
induced radiation till finite energy corrections become manifest
\cite{Neufeld:2010tz}. However, the multiplicity of medium-induced hadron
production is a
monotonously rising function of both pathlength and medium density.

\begin{figure}[htb]
\epsfig{file=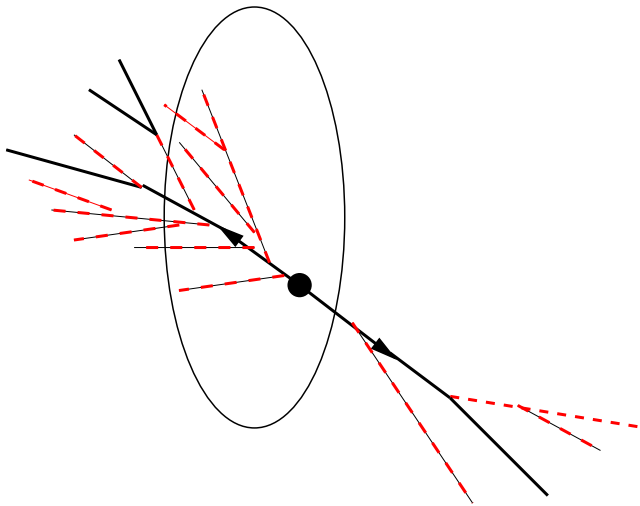, width=0.48\textwidth}
\caption{\label{F-Jets} Sketch of the hadron multiplicity asymmetry induced by
medium-modified showers: for a shower evolving in the medium, induced gluon
radiation leads to larger multiplicity production.}
\end{figure}

The basic idea how medium modified jets may generate an asymmetry in the charged
hadron multiplicity correlated out of the reaction plane is shown in Fig.~\ref{F-Jets}. 
Let us start with a random initial hard vertex position in the transverse plane:
if the vertex position is such that one jet is
close to the surface and escapes unmodified whereas the other jet propagates a
long distance through the medium, the average multiplicity of charged hadrons
for both jets is expected to be very different. Furthermore, this effect can be
estimated to be stronger in the up-down direction than in the left-right
direction, simply because the maximum variation in pathlength is larger for out
of plane emission than for in-plane emission.

Note that jets lead to an asymmetry in charged hadron production event by event
even without a medium modification, simply because the fragmentation of a hard
parton to a hadron jet is a probabilistic process, and also because jets in
hadronic collisions may be due to a pQCD subprocess like $qg \rightarrow qg$ in
which the fragmentation of the gluon in the out state is much softer from the
fragmentation of the quark. However, in the presence of a medium, the
medium-induced radiation is expected to dominate multiplicity production, and
the vacuum case has no mechanism to account for a correlation with the reaction plane. 
In the following, we use the in-medium shower evolution code YaJEM
(Yet another Jet Energy-loss Model) in the RAD (radiative energy loss) scenario
to estimate the jet-induced event by event charged hadron multiplicity
asymmetry.

\subsection{Model Description}

YaJEM is a Monte Carlo code to simulate the medium-modified evolution of a
parton shower. It is based on the PYSHOW algorithm
\cite{Bengtsson:1986hr,Bengtsson:1986et} for shower evolution in vacuum, to
which it reduces in the absence of a medium. A detailed description of the model
is given in \cite{Renk:2008pp,Renk:2009nz}, here we just summarize the essential
physics underlying the model.

In YaJEM, the evolution from the initial parton to a final state parton shower
is modelled as a series of branching processes $a \rightarrow b+c$ where $a$ is
called the parent parton and $b$ and $c$ are referred to as daughters. In QCD,
the allowed branching processes are $q \rightarrow qg$, $g \rightarrow gg$ and
$g \rightarrow q \overline{q}$.  The kinematics of a branching is described in
terms of the virtuality scale $Q^2$ and of the energy fraction $z$, where the
energy of daughter $b$ is given by $E_b = z E_a$ and of the daughter $c$ by $E_c
= (1-z) E_a$. It is convenient to introduce $t = \ln Q^2/\Lambda_{QCD}$ where
$\Lambda_{QCD}$ is the scale parameter of QCD. $t$ takes a role similar to a
time in the evolution equations, as it describes the evolution from some high
initial virtuality $Q_0$ ($t_0$) to a lower virtuality $Q_m$ ($t_m$) at which
the next branching occurs. In terms of the two variables, the differential
probability $dP_a$ for a parton $a$ to branch is
\begin{equation}
dP_a = \sum_{b,c} \frac{\alpha_s}{2\pi} P_{a\rightarrow bc}(z) dt dz
\end{equation}
where the splitting kernels $P_{a\rightarrow bc}(z)$ read
\begin{eqnarray}
&&P_{q\rightarrow qg}(z) = 4/3 \frac{1+z^2}{1-z} \label{E-qqg}\\
&&P_{g\rightarrow gg}(z) = 3 \frac{(1-z(1-z))^2}{z(1-z)}\label{E-ggg}\\
&&P_{g\rightarrow q\overline{q}}(z) = N_F/2 (z^2 + (1-z)^2)\label{E-gqq}.
\end{eqnarray}
We do not consider electromagnetic branchings. $N_F$ counts the number of active
quark flavours for given virtuality. The resulting system of equations
describing the branching processes in vacuum is solved numerically using MC
techniques utilizing the {\sc Pyshow} code
\cite{Bengtsson:1986hr,Bengtsson:1986et}.

In order to make the link from momentum space where the shower evolution takes
place to position space where the medium perturbations evolve, we assume that
the average formation time of a shower parton with virtuality $Q$ is developed
on the timescale $1/Q$, i.e. the average lifetime of a virtual parton with
virtuality $Q_b$ coming from a parent parton with virtuality $Q_a$ is in the
rest frame of the original hard collision (the local rest frame of the medium
may be different by a flow boost as the medium may not be static) given by
\begin{equation}
\label{E-Lifetime}
\langle \tau_b \rangle = \frac{E_b}{Q_b^2} - \frac{E_b}{Q_a^2}.
\end{equation}
We assume that the actual formation time can then be obtained from a probability
distribution
\begin{equation}
\label{E-RLifetime}
P(\tau_b) = \exp\left[- \frac{\tau_b}{\langle \tau_b \rangle}  \right]
\end{equation}
which we sample to determine the actual formation time of the fluctuation in
each branching.

The medium modification in the RAD scenario appears as a change of parton
virtuality $\Delta Q_a^2$ dependent on a medium parameter $\hat{q}$ where

\begin{equation}
\label{E-Qgain}
\Delta Q_a^2 = \int_{\tau_a^0}^{\tau_a^0 + \tau_a} d\zeta \hat{q}(\zeta).
\end{equation}

Here, the time $\tau_a$ is given by Eq.~(\ref{E-RLifetime}), the time $\tau_a^0$
is known in the simulation as the endpoint of the previous branching process and
the integration $d\zeta$ is along the eikonal trajectory of the
shower-initiating parton. If the parton is a gluon, the virtuality transfer from
the medium is increased by the ratio of their Casimir color factors,
$3/\frac{4}{3} =  2.25$.  $\hat{q}$ is a function of medium energy density
$\epsilon$ with the relation chosen such that the single pion suppression
observed in 200 AGeV central Au-Au collisions at RHIC is described
\cite{Renk:2009nz}. 

As a result of the medium-induced change in virtuality, additional gluon
radiation occurs which after hadronization leads to an increased multiplicity in
the shower.

\subsection{Results}

\begin{figure}[htb]
\resizebox{0.5\textwidth}{!}{
\centering  
\includegraphics{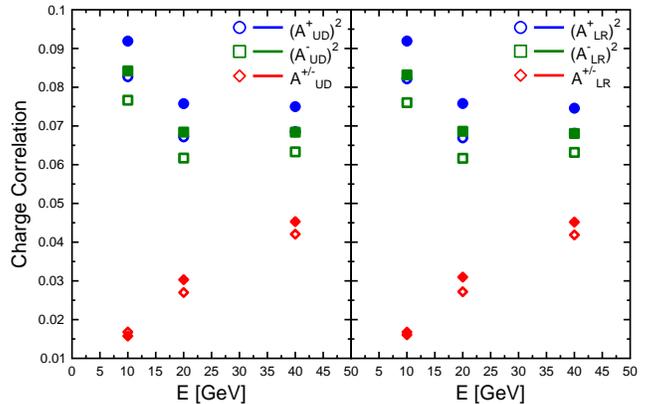}
}
\caption{(Color online) Contribution of medium-modified jets to the charged
particle asymmetry coefficients in the out-of-plane (left) and in-plane (right)
direction for three different initial parton energies (10,20 and 40 GeV) for
Au+Au collisions at $\sqrt{s_{\rm NN}}=200A$ GeV. Open symbols depict central
collisions ($b=2.4$ fm) while filled symbols represent mid-central collisions
($b=7.5$ fm.}
\label{fig_only_jet}      
\end{figure}

In order to estimate the event by event charged hadron asymmetry, we embed
back-to-back jet pairs of three different energies (10, 20 and 40 GeV per jet)
into a 3+1 d hydrodynamical simulation \cite{Nonaka:2006yn} on a random position
sampled from the binary collision profile. The jet pair then gets a random
orientation with respect to the reaction plane. We then propagate both jets
through the medium and evaluate Eq.~(\ref{E-Qgain}) to determine the medium
modification. After hadronizing the jets, we count the charged hadron content in
the four different quadrants and compute the asymmetry coefficients $(A^+)^2,
(A^-)^2$ and $A^{+-}$. The results are shown in Fig.~\ref{fig_only_jet}.

The result indicates that jets are in principle capable of creating substantial
charge correlations. In this scenario the charge correlation arises because of an asymmetry in the particle production. The correlation shown in Fig. \ref{fig_only_jet} shows that in a developing shower the balancing charges are preferably emitted in the same hemisphere, so positively and negativley charged particles tend to align. However, somewhat surprisingly, no pronounced difference
between up-down and left-right is visible. This can be traced back to a number
of reasons. First, Eq.~(\ref{E-Qgain}) evaluated in an expanding medium leads to
small modifications at late times, i.e. for long paths, as $\hat{q}$ becomes
very small when the energy density of the system is low. This means that in an
expanding medium, partons do not actually probe the full length difference
between in-plane and out-of-plane emission, but rather probe pathlength
difference during the time period when modifications are significant only, i.e.
the first $\sim$ 3-5 fm/c. Given this 'filter', the differences between in-plane
and out-of-plane emission are considerably smaller than estimated naively based
on geometry. Furthermore, especially for low $p_T$ hadron production, the
angular distribution of hadrons around the jet axis is wide
\cite{Renk:2009nz,Renk:2009hv}. Note that this can happen even if the jet shape
itself is rather narrow, as the jet shape traces the flow of momentum and not
multiplicity. The fact that there can be significant multiplicity created also
in
the down hemisphere even from a shower initiating parton travelling into the up
hemisphere further weakens the difference between in-plane and out-of-plane
emission.

\begin{figure}[h]
\resizebox{0.5\textwidth}{!}{
\centering  
\includegraphics{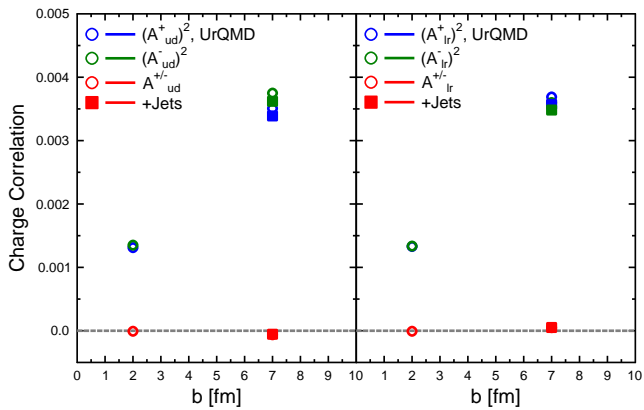}
}
\caption{(Color online) Calculation of the charged particle asymmetry
coefficients in the out-of-plane (left) and in-plane (right) direction for two
different centralities for Au+Au collisions at $\sqrt{s_{\rm NN}}=200A$ GeV. The
open circles depict the results from the hadronic transport approach (UrQMD)
that serves as a background for the full calculation including medium-modified
jets that is represented by full squares.}
\label{fig_jet_full}      
\end{figure}

So far, we have investigated the charge correlation coefficients created by a
single jet only. This is not a very realistic situation: First, high momentum
jets of e.g. 40 GeV are very rare events, so even if they create sizeable
asymmetries in charged multiplicity, such events may not have much weight in an
average over many events. Second, for low momentum jets, one cannot assume that
only a single jet per event contributes. However, it is immediately clear that
embedding multiple jets in an event weakens the asymmetry significantly. When we
account for these effects by mixing jets with the probability of a jet with
given energy being in the event determined from a LO pQCD calculation into a
UrQMD background \cite{Bass:1998ca,Bleicher:1999xi}, the picture changes
considerably: In Fig.~\ref{fig_jet_full}
the contribution of jets appears as an insignificant correction to the
background result. 

In hindsight, this could have been anticipated: While truly hard jets are too
rare to influence the result, low $p_T$ jets (below 5 GeV) tend to be created
frequently, but the presence of multiple jets in an event tends to cancel their
effect. Clearly, a bulk mechanism creating charge asymmetries is more efficient
than a rare probe.

Note also that it is apparent from Fig. \ref{fig_jet_full} that UrQMD background results show that a pure hadronic transport approach is not capable of translating initial state fluctuations to the final state distributions effectively enough to lead to a significant signal as the $A^{+/-}$ coefficients are on the order of $1 \cdot 10^{-6}$. 

\begin{figure}[h]
\resizebox{0.5\textwidth}{!}{
\centering  
\includegraphics{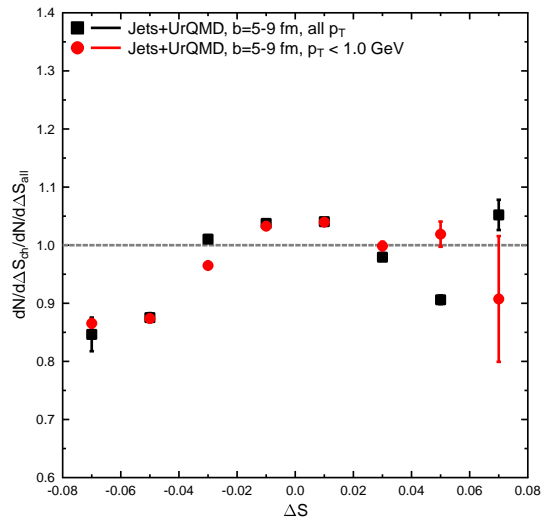}
}
\caption{(Color online) Calculation of the multi-particle correlation
coefficient in mid-central Au+Au collisions at $\sqrt{s_{\rm NN}}=200A$ GeV. The
full squares depict the results from the full calculation including medium-modified
jets to the hadronic transport approach (UrQMD) background event, whereas the full circles show the result only for particles with $p_{T}<1$ GeV.}
\label{fig_jet_phenix}      
\end{figure}
 
In Fig. \ref{fig_jet_phenix} the calculation of the multi-particle correlation coefficient is shown. Medium-modified jets, embedded in a UrQMD background lead to a convex shape of the ratio of the distributions in $\Delta S$. The result for jets alone leads to the same shape as the combined events. But again, the probability for jet production is very low and therefore the results resemble very much the UrQMD results that are shown in Fig.~\ref{fig_hybrid_phenix}. Furthermore, the influence of a $p_{T} < 1$GeV cut has been explored, since the preliminary measurements have been made in a limited $p_{T}$ range and the chiral magnetic effect is mostly expected in the bulk low transverse momentum region. The low $p_{T}$ result and the result for all charged particles are the same within error bars and especially the shape does not change. Since the distribution of the charge sensitive averaged sine differences has a higher peak around zero and lower flanks than the distribution that is not charge specific this result indicates that particles with opposite charges are preferably emitted in the same direction than in different directions.  The underlying physics mechanism in both cases is correlated particle production from 'clusters' - in one case jets, in the other decaying resonances.

\section {Initial State Fluctuations}
\label{granularity}
\subsection{Overview}
In the second part of this article the charged particle asymmetry that arises
from well-known bulk medium effects is investigated. The basic idea is that
energy density clusters produced by the initial binary scatterings in heavy
ion reactions might lead to asymmetry correlations in the particle production
in different hemispheres. Due to the development of radial flow during the
medium evolution the initial coordinate space granularity is translated to a
resulting momentum space anisotropy. These clusters are expected to enhance the
multiplicity in a specific hemisphere which results in alignment of the charged
particle production instead of separation. In this picture, resonance decays might explicitly introduce a charge correlation of the outgoing particles, if e.g. a $\rho$ meson decays into $\pi^+ +\pi^-$ and there is enough collective flow, both of the pions will preferably be have a momentum in a similar direction. 

S. Pratt recently estimated the effects of elliptic flow, momentum and charge
conservation on the observable charged particle correlation measurement
\cite{Pratt:2010gy}. Our approach follows a similar spirit but the effect of
initial state fluctuations is emphasized in addition. This kind of
event-by-event fluctuations has been found to be crucial also for other
two-particle correlation measurements (the 'ridge') at the
highest RHIC energies \cite{Takahashi:2009na}. 

\subsection{Model Description}
The charged particle asymmetry that is generated by clusters in the initial
energy density distribution is quantified by using a hybrid approach based on
the UrQMD hadronic transport approach including a
(3+1)-dimensional 
one fluid ideal hydrodynamic evolution
\cite{Rischke:1995ir,Rischke:1995mt} for the hot and dense stage of the reaction
while the early non-equilibrium stage and the final decoupling is treated in the
hadronic cascade \cite{Petersen:2008dd,Petersen:2009vx} \footnote{The code is
available as UrQMD-3.3p1 at http://urqmd.org}.   

The event-by-event-setup that is naturally employed in this approach is crucial
for the observation of charged particle asymmetries. For Au+Au collisions at the
highest RHIC energy the starting time for the hydrodynamic evolution has been
chosen to be $t_{\rm start}=0.3-1.8$ fm depending on the amount of fluctuations
to fit the final pion multiplicity at midrapidity, since this is crucial for the present study and allows us to investigate the effect of the initial state granularity on the charge correlations. How the change in the starting time affects other bulk observables like spectra and elliptic flow will be discussed in a further publication \cite{Petersen:2010zt}. Only the matter around
midrapidity
($|y|<2$) is considered to be locally thermalized and takes part in the ideal
hydrodynamic evolution. The more dilute spectator/corona regions are treated in
the hadronic cascade approach. To map the point particles from the UrQMD initial
state to energy,
momentum and net baryon density distributions three-dimensional Gaussian
distributions are used that represent one particle each
\cite{Steinheimer:2007iy}. 

\begin{equation}
\label{eq_gauss}
\footnotesize{\epsilon_{\rm cf}(x,y,z)=N  \exp{-\frac{(x-x_{p})^2+(y-y_{p})^2
+(\gamma_z(z-z_{p}))^2}{2 \sigma_{\rm Gauss}^2}}}\quad,
\end{equation}
where $N=(\frac{1}{2 \pi })^{\frac{3}{2}} \frac{\gamma_z}{\sigma_{\rm Gauss}^3} 
E_{\rm cf} $ provides the proper normalisation,
$\epsilon_{\rm cf} $ and $E_{\rm cf} $ are the energy density and total energy
of the 
particle in the computational frame, while $(x_p,y_p,z_p)$ is the position
vector of the particle.
Summing over all single particle distribution functions leads to distributions
of energy-, momentum- and baryon number-densities in each cell.

The transition from the hydrodynamic evolution to the transport approach when
the matter
is diluted in the late stage is treated as a gradual transition on an
approximated iso-eigentime hyper-surface (see \cite{Li:2008qm,Steinheimer:2009nn} for details).
The final rescatterings and resonance decays are taken into account in the
hadronic cascade.

To investigate the effect of different initial state granularities the width of
the Gaussian distribution (\ref{eq_gauss}) is varied from
$\sigma_{\rm Gauss}=0.8-2.0$ fm. Please note that this is still an event-by-event
setup and in this respect different from averging over many fluctuating initial
conditions as has been done in \cite{Petersen:2010md}. In a hydrodynamic calculation starting from smooth initial profiles the correlation observable would  be zero, since there are no event-by-event fluctuations included.  
 
\begin{figure}[h]
\resizebox{0.6\textwidth}{!}{
\centering  
\includegraphics{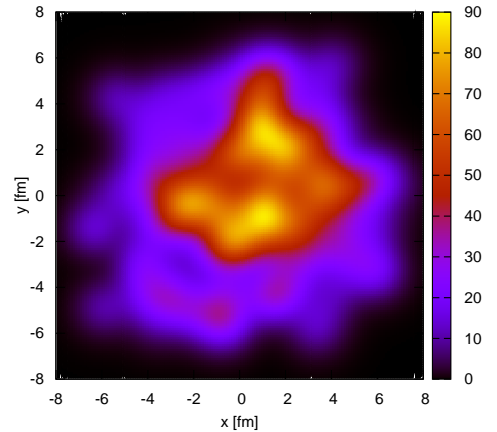}
}
\caption{(Color online) Energy density distribution in the transverse plane for
a central (b=0 fm) Au+Au collision at $\sqrt{s_{\rm NN}}=200A$ GeV with
$\sigma_{\rm Gauss}=0.8$ fm and $t_{\rm start}=0.3$ fm.} 
\label{fig_ini_sig0.8}      
\end{figure}

\begin{figure}[h]
\resizebox{0.6\textwidth}{!}{
\centering  
\includegraphics{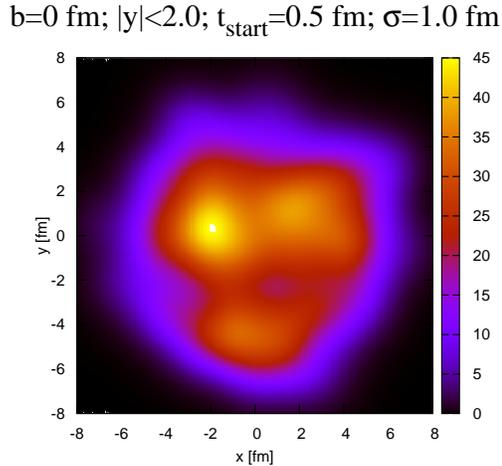}
}
\caption{(Color online) Energy density distribution in the transverse plane for
a central (b=0 fm) Au+Au collision at $\sqrt{s_{\rm NN}}=200A$ GeV with
$\sigma_{\rm Gauss}=1.0$ fm and $t_{\rm start}=0.5$ fm.}
\label{fig_ini_sig1.0}      
\end{figure}

\begin{figure}[h]
\resizebox{0.6\textwidth}{!}{
\centering  
\includegraphics{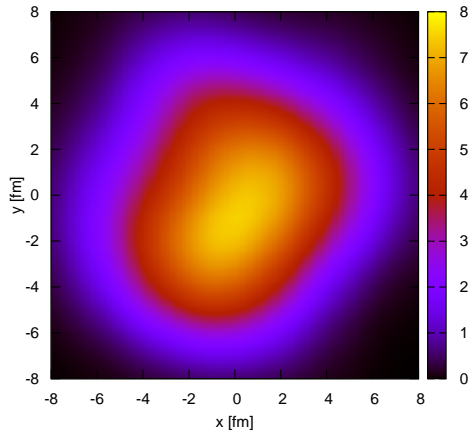}
}
\caption{(Color online) Energy density distribution in the transverse plane for
a central (b=0 fm) Au+Au collision at $\sqrt{s_{\rm NN}}=200A$ GeV with
$\sigma_{\rm Gauss}=2.0$ fm and $t_{\rm start}=1.8$ fm.}
\label{fig_ini_sig2.0}      
\end{figure}

Fig. \ref{fig_ini_sig0.8}-\ref{fig_ini_sig2.0} show the initial energy density
distribution as it is obtained from UrQMD for three different values of the
Gaussian width $\sigma_{\rm Gauss}=0.8$, 1.0 and 2.0 fm. A smaller width leads
to entropy production during the ideal hydrodynamic evolution indicating that
the solution of the differential equation is not stable anymore. Therefore, the
highest fluctuations are generated with the smallest width of 0.8 fm, while for
larger widths the initial fluctuations are more and more smeared out. Smaller
characteristic sizes could be physically motivated by a color flux tube picture
where structures in the transverse plane on the order of $0.3$ fm would be
expected. Our parameter choices stay in a regime where the implementation is
technically stable and the characteristic sizes correspond to typical sizes of a
proton.   

\subsection{Results}

\begin{figure}[h]
\resizebox{0.5\textwidth}{!}{
\centering  
\includegraphics{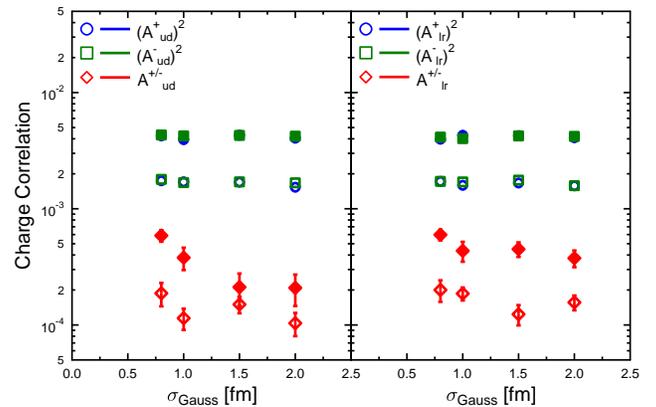}
}
\caption{(Color online) Calculation of the charged particle asymmetry
coefficients in the out-of-plane (left) and in-plane (right) direction for four
different values of $\sigma_{\rm Gauss}$. The open symbols depict the results
for central ($b<3.4$ fm) Au+Au collisions at $\sqrt{s_{\rm NN}}=200A$ GeV and
the filled symbols represent mid-central ($b=5-9$ fm) Au+Au collisions.}
\label{fig_nch_corr_sigdep}      
\end{figure}

The hybrid approach provides the opportunity to study the influence of initial
state fluctuations that can be varied in size without adjusting any other
parameter on event-by-event particle production in different hemispheres.
Initial state clusters and a coordinate space -momentum space correlation due to
radial flow leads to observable differences in the charged particle production
in different parts of the system (see Fig. \ref{fig_nch_corr_sigdep}). The UD
and LR charged particle asymmetry coefficients have been calculated for four
different initial granularities in the hybrid approach. The open symbols depict
the results for central ($b<3.4$ fm) Au+Au collisions at $\sqrt{s_{\rm
NN}}=200A$ GeV and the filled symbols represent mid-central ($b=5-9$ fm) Au+Au
collisions. The overall fluctuations of charged particles are larger in
non-central collisions than in central collisions because the overall
multiplicity is smaller. The opposite charge asymmetry is positive and similar
in the out-of-plane and the in-plane direction. This confirms the expectations
that in this scenario the particles tend to align independent of their charge.
Within our limited statistics, the absolute value of the difference between UD and LR can be constrained to
be smaller than $1\cdot 10^{-3}$.

Higher granularity in the initial state leads to higher opposite charge
asymmetry coefficients that are on the order of the preliminary STAR data \cite{STAR_Talk}. This
physics mechanism will contribute to the background of a potential measurement
of the chiral magnetic effect and needs to be taken into account, before trying
to look for a potentially much smaller signal. On the other hand, the comparison of
hybrid calculation with different initial state granularities to event-by-event
correlation and fluctuation measurements might be helpful to constrain the
structure of the initial state of relativistic heavy ion reactions. 

\begin{figure}[h]
\resizebox{0.5\textwidth}{!}{
\centering  
\includegraphics{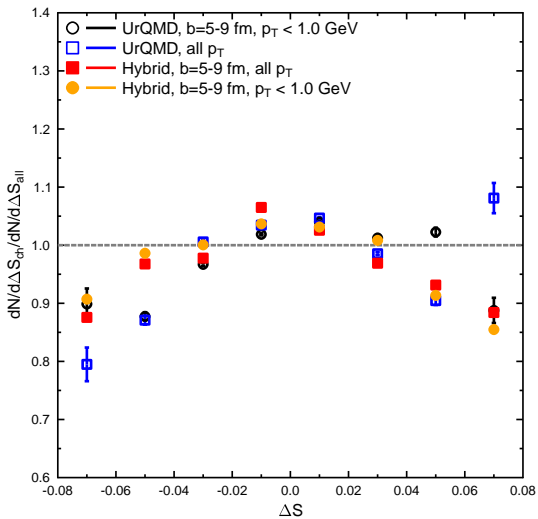}
}
\caption{(Color online) Calculation of the multi-particle correlation
coefficient in mid-central Au+Au collisions at $\sqrt{s_{\rm NN}}=200A$ GeV. The open symbols 
depict the results from the hadronic transport approach (UrQMD) (circles indicate an additional $p_{T}<1$ GeV cut) and the filled symbols represent calculations within the hybrid approach.}
\label{fig_hybrid_phenix}      
\end{figure}
 
Fig. \ref{fig_hybrid_phenix} shows the result for the multi-particle correlation coefficient that has been defined in Section \ref{sec_phenix_def} from the UrQMD hadronic transport approach (open symbols) and the hybrid approach (filled symbols). It has been checked that the results for this observable do not depend on the choice of $\sigma_{\rm Gauss}$. The low transverse momentum region shows the same behaviour as without $p_{T}$ cut. If one reduces the number of particles by acceptance (or other) cuts, the distribution broadens, however does not change qualitatively in shape. Again the correlations induced by these approaches lead to a convex shape and can be attributed to resonance decays whereas a charge separation scenario as expected from the chiral magnetic effect leads to a concave shape in this observable. The initial state fluctuations have to be considered as a non-trivial source of correlations of the particles emitted in a heavy ion reaction.

It is likely that any measurement of the observable is dominated by resonance decay effects which are known to be present in nature. In order to observe a different mechanism (like the CME) at work, this distribution has to be subtracted. In order to get an idea of the sensitivity of the results to the subtraction, we have tried to isolate the effect of initial state fluctuations by subtracting the UrQMD distributions in $\Delta S$ from the hybrid model distributions. It turns out that the procedure is very sensitive to the subtraction - we were able to observe even qualitative differences in shape which could potentially mock up the CME signal.

\section{Summary and Conclusions}
\label{summary}
We have investigated charge asymmetry and multi-particle correlations  with the reaction plane 
generated in relativistic heavy-ion collisions at RHIC within a realistic transport approach 
containing a full medium evolution and medium-modified jets. While jets are in principle
capable of generating the aforementioned multi-particle correlations, we were able to rule them
out as a significant contributor to such correlations due to the low jet production rate.
However, we find that initial state fluctuations can generate values for the charge 
asymmetry coefficients on the same order of magnitude as seen by experiment.
Our results were obtained within a hybrid transport approach, based on an event-by-event
hydrodynamic evolution, with fluctuating initial conditions and microscopic hadronic afterburner
for the realistic treatment of hadronic final state effects, including resonance decays. 
The precision of our calculation  was not sufficient to quantify the difference between in-plane and out-of-plane, due to the very large computational effort involved.
However, the coefficients themselves show no sign of charge separation.
The multi-particle correlation measurement proposed by the PHENIX collaboration appears to 
be potentially the most suitable of the studied observables for differentiating a CME 
from other mechanisms. 
However, due to the need of subtracting known background effects like resonance decays, it is 
currently unclear  whether the predicted concave shape as a signal for the CME can be robustly established.  

\section{Acknowledgements}
\label{ack}
We are grateful to the Open Science Grid and the Center for the Scientific
Computing (CSC) at Frankfurt
for the computing resources. The authors thank Fuqiang Wang, Roy Lacey and Berndt M\"uller for helpful discussions and Dirk Rischke for providing the 1
fluid hydrodynamics code. H.P. acknowledges a Feodor Lynen fellowship of
the Alexander von Humboldt foundation. This work was supported in part by U.S.
department of
Energy grant DE-FG02-05ER41367 and by project 133005 of the Academy of Finland. T.R. is supported by an Academy Research Fellowship of the Academy of Finland. 



\begin{thebibliography}{99}

\bibitem{Arsene:2004fa}
Arsene I {\em et~al.\/} (BRAHMS) 2005 {\em Nucl. Phys.\/} {\bf A757} 1--27

\bibitem{Adcox:2004mh}
Adcox K {\em et~al.\/} (PHENIX) 2005 {\em Nucl. Phys.\/} {\bf A757} 184--283

\bibitem{Back:2004je}
Back B~B {\em et~al.\/} 2005 {\em Nucl. Phys.\/} {\bf A757} 28--101

\bibitem{Adams:2005dq}
Adams J {\em et~al.\/} (STAR) 2005 {\em Nucl. Phys.\/} {\bf A757} 102--183

\bibitem{Muller:2006ee}
Muller B and Nagle J~L 2006 {\em Ann. Rev. Nucl. Part. Sci.\/} {\bf 56} 93--135

\bibitem{Gyulassy:2004zy}
Gyulassy M and McLerran L 2005 {\em Nucl. Phys.\/} {\bf A750} 30--63

\bibitem{Kharzeev:2007jp}
  D.~E.~Kharzeev, L.~D.~McLerran and H.~J.~Warringa,
  Nucl.\ Phys.\  A {\bf 803}, 227 (2008)

\bibitem{Fukushima:2008xe}
  K.~Fukushima, D.~E.~Kharzeev and H.~J.~Warringa,
  Phys.\ Rev.\  D {\bf 78}, 074033 (2008)
  
\bibitem{Fukushima:2010vw}
  K.~Fukushima, D.~E.~Kharzeev and H.~J.~Warringa,
  Phys.\ Rev.\ Lett.\  {\bf 104}, 212001 (2010)
  
\bibitem{Asakawa:2010bu}
  M.~Asakawa, A.~Majumder and B.~Muller,
  Phys.\ Rev.\  C {\bf 81}, 064912 (2010)

 
  
\bibitem{Abelev:2009uh}
  B.~I.~Abelev {\it et al.}  [STAR Collaboration],
  Phys.\ Rev.\ Lett.\  {\bf 103}, 251601 (2009)
  
\bibitem{Schlichting:2010na}
  S.~Schlichting and S.~Pratt,
  arXiv:1005.5341 [nucl-th].

\bibitem{Pratt:2010gy}
  S.~Pratt,
  arXiv:1002.1758 [nucl-th].
 
\bibitem{Wang:2009kd}
  F.~Wang,
  Phys.\ Rev.\  C {\bf 81}, 064902 (2010)

 \bibitem{Bzdak:2009fc}
  A.~Bzdak, V.~Koch and J.~Liao,
  Phys.\ Rev.\  C {\bf 81}, 031901 (2010)

\bibitem{STAR_Talk}
F.~Wang, talk given at 'P- and CP- odd Effects in Hot and Dense Matter' on behalf of the STAR collaboration, April 2010.  

\bibitem{PHENIX_Talk}
N.~Ajitanand, talk given at the annual AGS-RHIC user meeting 2010 on behalf of the PHENIX collaboration.
  
\bibitem{Abelev:2009txa}
 B.~I.~Abelev {\it et al.}  [STAR Collaboration],
  Phys.\ Rev.\  C {\bf 81}, 054908 (2010)



\bibitem{Baier:1996sk}
  R.~Baier, Y.~L.~Dokshitzer, A.~H.~Mueller, S.~Peigne and D.~Schiff,
  Nucl.\ Phys.\  B {\bf 484}, 265 (1997)

\bibitem{Zakharov:1997uu}
  B.~G.~Zakharov,
  JETP Lett.\  {\bf 65}, 615 (1997)

\bibitem{Neufeld:2010tz}
  R.~B.~Neufeld and T.~Renk,
  Phys.\ Rev.\  C {\bf 82}, 044903 (2010)

\bibitem{Bengtsson:1986hr}
  M.~Bengtsson and T.~Sjostrand,
  Phys.\ Lett.\  B {\bf 185}, 435 (1987).

\bibitem{Bengtsson:1986et}
  M.~Bengtsson and T.~Sjostrand,
  Nucl.\ Phys.\  B {\bf 289}, 810 (1987).

\bibitem{Renk:2008pp}
  T.~Renk,
  Phys.\ Rev.\  C {\bf 78}, 034908 (2008)

\bibitem{Renk:2009nz}
  T.~Renk,
  Phys.\ Rev.\  C {\bf 79}, 054906 (2009)

\bibitem{Nonaka:2006yn}
  C.~Nonaka and S.~A.~Bass,
  Phys.\ Rev.\  C {\bf 75}, 014902 (2007)

\bibitem{Renk:2009hv}
  T.~Renk,
  Phys.\ Rev.\  C {\bf 80}, 044904 (2009)

\bibitem{Bass:1998ca}
  S.~A.~Bass {\it et al.},
  Prog.\ Part.\ Nucl.\ Phys.\  {\bf 41}, 255 (1998)
  [Prog.\ Part.\ Nucl.\ Phys.\  {\bf 41}, 225 (1998)]

\bibitem{Bleicher:1999xi}
  M.~Bleicher {\it et al.},
  J.\ Phys.\ G {\bf 25}, 1859 (1999)

\bibitem{Takahashi:2009na}
  J.~Takahashi, B.~M.~Tavares, W.~L.~Qian, F.~Grassi, Y.~Hama, T.~Kodama and
N.~Xu,
  Phys.\ Rev.\ Lett.\  {\bf 103}, 242301 (2009)

\bibitem{Rischke:1995ir}
  D.~H.~Rischke, S.~Bernard and J.~A.~Maruhn,
  Nucl.\ Phys.\  A {\bf 595}, 346 (1995)

\bibitem{Rischke:1995mt}
  D.~H.~Rischke, Y.~Pursun and J.~A.~Maruhn,
  Nucl.\ Phys.\  A {\bf 595}, 383 (1995)
  [Erratum-ibid.\  A {\bf 596}, 717 (1996)]

\bibitem{Petersen:2008dd}
  H.~Petersen, J.~Steinheimer, G.~Burau, M.~Bleicher and H.~Stocker,
  Phys.\ Rev.\  C {\bf 78}, 044901 (2008)

\bibitem{Petersen:2009vx}
  H.~Petersen and M.~Bleicher,
  Phys.\ Rev.\  C {\bf 79}, 054904 (2009)

\bibitem{Petersen:2010zt}
  H.~Petersen, C.~Coleman-Smith, S.~A.~Bass and R.~Wolpert,
  arXiv:1012.4629 [nucl-th].

\bibitem{Steinheimer:2007iy}
  J.~Steinheimer, M.~Bleicher, H.~Petersen, S.~Schramm, H.~Stocker and
D.~Zschiesche,
  Phys.\ Rev.\  C {\bf 77}, 034901 (2008)

\bibitem{Li:2008qm}
  Q.~f.~Li, J.~Steinheimer, H.~Petersen, M.~Bleicher and H.~Stocker,
  Phys.\ Lett.\  B {\bf 674}, 111 (2009)

\bibitem{Steinheimer:2009nn}
  J.~Steinheimer, V.~Dexheimer, H.~Petersen, M.~Bleicher, S.~Schramm and H.~Stoecker,
 Phys.\ Rev.\ C {\bf 81}, 044913 (2010) 


\bibitem{Petersen:2010md}
  H.~Petersen and M.~Bleicher,
  Phys.\ Rev.\  C {\bf 81}, 044906 (2010)


\end{thebibliography}
\end{document}